\newcommand{\nin}{\noindent}

\newcommand{\non}{\nonumber}

\newtheorem{@theorem}{Theorem}
\newcommand{\theorem}[1]{\begin{@theorem}#1\end{@theorem}}

\newtheorem{@satz}{Satz}
\newcommand{\satz}[1]{\begin{@satz}#1\end{@satz}}

\newtheorem{@claim}{Claim}
\newcommand{\claim}[1]{\begin{@claim}#1\end{@claim}}

\newtheorem{@behauptung}{Behauptung}
\newcommand{\behauptung}[1]{\begin{@behauptung}#1\end{@behauptung}}

\newtheorem{@proposition}{Proposition}
\newcommand{\proposition}[1]{\begin{@proposition}#1\end{@proposition}}

\newtheorem{@lemma}{Lemma}
\newcommand{\lemma}[1]{\begin{@lemma}#1\end{@lemma}}

\newtheorem{@corollary}{Corollary}
\newcommand{\corollary}[1]{\begin{@corollary}#1\end{@corollary}}

\newtheorem{@korollar}{Korollar}
\newcommand{\korollar}[1]{\begin{@korollar}#1\end{@korollar}}

\newtheorem{@definition}{Definition}
\newcommand{\definition}[1]{\begin{@definition}#1\end{@definition}}

\newtheorem{@assumption}{Assumption}
\newcommand{\assumption}[1]{\begin{@assumption}#1\end{@assumption}}

\newtheorem{@annahme}{Annahme}
\newcommand{\annahme}[1]{\begin{@annahme}#1\end{@annahme}}

\newtheorem{@remark}{Remark}
\newcommand{\remark}[1]{\begin{@remark}#1\end{@remark}}

\newtheorem{@bemerkung}{Bemerkung}
\newcommand{\bemerkung}[1]{\begin{@bemerkung}#1\end{@bemerkung}}

\newtheorem{@example}{Example}
\newcommand{\example}[1]{\begin{@example}#1\end{@example}}

\newtheorem{@beispiel}{Beispiel}
\newcommand{\beispiel}[1]{\begin{@beispiel}#1\end{@beispiel}}

\newcommand{\C}{\hbox{C\hskip-0.5em\lower-0.1ex\hbox{\vrule
                      height1.34ex width0.07em }}\hskip0.50em}

\newcommand{\G}{\hbox{G\hskip-0.525em\lower-0.081ex\hbox{\vrule
                      height1.4ex width0.07em }}\hskip0.50em}

\renewcommand{\O}{\hbox{O\hskip-0.525em\lower-0.095ex\hbox{\vrule
                      height1.45ex width0.07em}}\hskip0.50em}

\newcommand{\Q}{\hbox{Q\hskip-0.525em\lower-0.097ex\hbox{\vrule
                      height1.47ex width0.07em}}\hskip0.50em}

\newcommand{\U}{\hbox{U\hskip-0.45em\lower-0.02ex\hbox{\vrule
                height1.54ex width0.07em}}\hskip0.50em}

\newcommand{\1}{1\hskip-0.28em \text{I}}

\newcommand{\argmin}{\mbox{argmin}}

\documentclass[a4paper,12pt]{article}

\usepackage[utf8]{inputenc}
\usepackage[]{amstext}
\usepackage[]{color}
\usepackage[]{amsmath}
\usepackage[]{amssymb}
\usepackage{multirow}
\usepackage[]{epsfig}
\usepackage[]{graphicx}
\usepackage[]{lscape}
\usepackage{ulem}
\usepackage[margin=1.25in]{geometry}
\usepackage[]{bm}
\usepackage{mathrsfs}
\usepackage{lipsum}
\usepackage{ragged2e}
\usepackage{courier}
\usepackage{adjustbox}
\usepackage{setspace}
\usepackage{pdflscape}
\usepackage{amsmath} 
\usepackage{lipsum}  
\usepackage{qtree}
\usepackage{rotating}
\usepackage{amssymb}
\usepackage{bbold}
\usepackage{dsfont}
\usepackage{xurl}
\usepackage{array}
\usepackage[linguistics]{forest}
\usepackage{tikz}
\usepackage{booktabs}
\usepackage{multirow}
\usepackage{placeins}
\usepackage{longtable}
\usepackage{threeparttable}

\usepackage{subcaption}
\usepackage[]{graphicx}
\usepackage{threeparttablex}
\usepackage{rotating}
\usepackage{tabularx}
\usepackage[]{lscape}
\usepackage{xcolor}
\usepackage{ulem}
\usepackage{ltablex}
\usepackage{appendix}

 \usepackage[utf8]{inputenc}
\usepackage{listings}
\usepackage{amsmath,blkarray,booktabs, bigstrut}

\usetikzlibrary{arrows}
\usetikzlibrary{shapes}

\begin{document}

\title{On the implications of proportional hazards assumptions for competing risks modelling}

\author{
Simon M.S. Lo\footnote{United Arab Emirates University, Department of Economics and Finance, E--mail: losimonms@yahoo.com.hk} \\
Ralf A. Wilke\footnote{Copenhagen Business School, Department of Economics, Porcel{\ae}nshaven 16A, 2000 Frederiksberg, DK, E--mail: rw.eco@cbs.dk}\\
Takeshi Emura\footnote{University of Hiroshima, E--mail: takeshiemura@gmail.com}
}
\maketitle
\thispagestyle{empty}

\begin{abstract}
The assumption of hazard rates being proportional in covariates is widely made in empirical research and extensive research has been done to develop tests of its validity. This paper does not contribute on this end. Instead, it gives new insights on the implications of proportional hazards (PH) modelling in competing risks models. It is shown that the use of a PH model for the cause-specific hazards or subdistribution hazards can strongly restrict the class of copulas and marginal hazards for being compatible with a competing risks model. The empirical researcher should be aware that working with these models can be so restrictive that only degenerate or independent risks models are compatible. Numerical results confirm that estimates of cause-specific hazards models are not informative about patterns in the data generating process.\\
\nin Keywords: dependent competing risks, copula, identifiability
\end{abstract}

\section{Introduction}
The hazard function in the single risk Cox proportional hazards (PH) regression model is $\lambda(t|z) = \lambda_0(t)\psi(z)$. Its multiplicative separability of the baseline hazard $\lambda_0(t)$ and the covariate function $\psi(z)$ has important practical advantages such $\psi(z)$ can be estimated by means of partial likelihood methods without specifying $\lambda_0(t)$. It seems natural to carry this type of proportionality restriction over to the competing risks model with more than one failure type. Examples are the Cox-type cause-specific proportional hazard (CSH) model (Cox, 1972, Ozenne et al., 2017) and the subdistribution proportional hazards (SDH) model (Fine and Gray, 1999). These models are popular among practitioners as the CSH and SDH (in contrast to the marginal hazards) are identifiable without restrictions on the risk dependence and share the same numerical advantages of the Cox PH model in the single-risk setting. The widespread use of these models (Google scholar citation counts for Fine and Gray: $\sim$14,500) justifies a deeper analysis of the implications of such assumptions on other model components which are part of the data generating process: the marginal hazards  $\lambda_j(t|z)$ for risk $j$ and the dependence structure between competing risks (copula). This paper shows that imposing proportionality restrictions on the CSHs or SDHs in the competing risks model is not as natural and general as commonly thought. Assuming separability of the CSH or SDH in $t$ and $z$ can imply crucial restrictions on $\lambda_j(t|z)$ and the copula. Testing the validity of the \textit{PH property} has been extensively studied in the literature (e.g. Grambsch and Therneau, 1994; Schoenfeld, 1980) and is a relevant problem in various disciplines such medicine (Abeysekera and Sooriyarachchi, 2009), engineering (Kumar and Klefsj\"{o}, 1994) and social sciences (Keele, 2010). The point of the paper is not about that proportional hazards restrictions may not be supported or valid. It is about showing that PH restrictions on the CSH or SDH can heavily restrict the data generating process of the competing risks model. In some cases they are so strong that they imply the model is degenerate or must have independent risks. This observation is important for applied research, because it shows that a model for te CSH or SDH does not leave the marginal hazards and copula unrestricted. This is despite that they are ignored in the modelling and treated as a black box. Our results suggest that disregarding the implications for the data generating process is risky as it can lead to unrealistic model restrictions.

We show that the PH assumption for the CSH and the SDH models can only be theoretically justified under certain restrictions on the marginal hazards. In the case of the Gumbel copula, for which we obtain a closed form solution for the marginal hazards, these must be proportional across risks (the \textit{risk-proportionality property}, RP). Under RP, not only the covariates induce a proportional shift of the  marginal hazards, i.e. $\lambda_{j}(t|z_1) \propto \lambda_j(t|z_2)$ for any two $z_1$ and $z_2$ for the $j$'th risk, but also the marginal hazards for two risks are proportional, e.g., $\lambda_{1}(t|z) \propto \lambda_2(t|z)$ for any $t$ and $z$ for risks 1 and 2. The implications of RP were studied by Kalbfleisch and Prentice (2002), and further explored by Lo et al. (2024). In this paper, we show that RP simplifies the competing risks model, which in turn induces identifiability of the marginal hazards and the degree of risk dependence. It is common to use numerically convenient partial likelihood methods for inference regarding the covariate effects on the CSHs. By establishing the link to the marginal hazards, partial likelihood methods can be also used for inference regarding the marginal hazards. The PH assumption on the SDH is even more restrictive as it additionally implies a Fr\'{e}chet-Hoeffding upper bound copula for the competing risks dependence. It can be also shown that other popular copulas such as Clayton or Frank are incompatible with Cox models for the CSH and SDH. This highlights that PH assumptions on the CSH and SDH can substantially restrict the valid class of copulas and the marginal distributions of the underlying data generating process.

The rest of the paper is organised as follows. Section \ref{sec:model} introduces the model and the theoretical results. Section \ref{sec:est} presents estimation by partial likelihood. Section \ref{sec:sim} reports simulation results to documents the finite sample performance and convergence behaviour. This is followed by an application to employment duration data in Section \ref{sec:appl}

\section{The Model \label{sec:model}}
We consider a competing risks model with two dependent risks. An extension to a model with more than two risks is straightforward but inevitably complicates the notations. In the case the dependence structure between risks is an Archimedean copula, risk-pooling methods can be easily applied that reduce any multiple risks model to a 2 risks model (Lo and Wilke, 2010). The duration of risk 1 is $T_1$ and the duration of risk 2 is $T_2$. $Z$ is a vector of covariates affecting both durations. The competing risks model is generated by a copula, denoted as $C(s_1,s_2;\theta)$ with unknown parameters $\theta$, and two marginal survival functions $S_j(t|z;\beta_j)= \exp(-\Lambda_j(t|z;\beta_j))$ for risks $j=1,2$. The marginal hazard
$$\lambda_j(t|z;\beta_j) = \text{lim}_{dt\rightarrow 0} \Pr(t\leq T_j< t+dt | T_j>t, z; \beta_j)$$
\nin has risk-specific parameters $\beta_j$. The marginal cumulative hazard is $\Lambda_j(t|z;\beta_j) = \int_0^{t}\lambda_j(u|z;\beta_j)du$.  The joint survival function of $T_1$ and $T_2$ conditional on $Z$ is $J(t_1,t_2|z) = \Pr(T_1>t_1, T_2>t_2|z)$. In the copula model it is formed by
\begin{eqnarray}
J(t_1,t_2|z; \theta, \beta_1, \beta_2) &=& C\big(S_1(t_1|z;\beta_1), S_2(t_2|z;\beta_2); \theta\big). \label{model0}
\end{eqnarray}
\nin According to the Sklar's theorem, there is a unique $C$ corresponding to any combination of $J$, $S_1$, and $S_2$. There is therefore no loss of generality by adopting a copula model for the risk dependence. We define $S(t|z)$ as the overall survival, which is the joint distribution in (\ref{model0}) when $t_1=t_2=t$, such that $S(t|z) = J(t,t|z)=C(S_1(t|z),S_2(t|z))$. While at this point, we do not restrict the family of copulas, it is assumed that the copula does not depend on $z$. Since risks 1 and 2 are competing risks, we do not observe $T_1$ and $T_2$ but only the minimum $T=\min\{T_1,T_2\}$ and the risk indicator $\delta =\argmin_j\{T_j\}$, $j\in\{1,2\}$.

\subsection{Structural versus a reduced-form modelling of competing risks}
The copula-based competing risks model in (\ref{model0}) can be considered as a \textit{structural model} as it describes the underlying data generating process of the observed $(T,\delta|Z)$. It is in contrast to the \textit{reduced-form model}, which we will introduce later. Correspondingly, $(\theta, \beta_1, \beta_2)$ are the \textit{structural parameters} that describe the data generating process completely. Our goal is to identify these structural parameters to identify the marginal survivals $S_j$ for $j=1,2$ and the partial effects of the elements of $Z$ on $(T_1, T_2)$ and the final outcomes  $(T,\delta|Z)$. In particular, It allows us to investigate how a change in the $k$th covariate $Z_k$ affect durations $T_1$ and $T_2$ differently. These are the so-called causal effects which are normally of interest in applied work. Take medical research as an example. Practitioners are often interested in knowing the effect of a medical treatment on the duration of relapse.
It is not possible to identify the causal effect on relapse $T_1$ by modelling $T$, as $T_1$ is censored by $T_2$, e.g., death due to other diseases as a side effect of the treatment. We can only isolate and evaluate the effectiveness of this treatment on $T_1$  using the structural model.  Another merit of the structural approach is that it allows us to conduct counterfactual analysis, such as how outcomes would change in response to changes in the structural parameters. For instance, what would be the probability of relapse, when a new medicine affects relapses and side-effects differently. In the rest of the paper, we use the terminology \textit{causal effect} in relation to the structural model. In contrast, the reduced form model only describes observational patterns.

We state the two approaches in terms of their likelihood functions to formalise their conceptional differences. Let $f_j(t|z)$ be the incidence function or so-called sub-density function, defined as
$$f_j(t|z)  = \text{lim}_{dt\rightarrow 0} \Pr(t\leq T< t+dt,\delta=j| z).$$
It is the probability density function for the event that a failure of risk $j$ is observed at $T=t$. A generic likelihood function for the competing risks model is
\begin{eqnarray}
L_i(\Sigma |t_i,z_i) = \big[f_1(t_i|z_i;\Sigma) \big]^{2-\delta_i} \big[f_2(t_i|z_i;\Sigma)\big]^{\delta_i -1}, \label{like1}
\end{eqnarray}
\nin for observations $i= 1,\ldots,n$, where $\Sigma$ is a vector of unknown parameters involved in $f_1$ and $f_2$. The likelihood function in (\ref{like1}) is generic, as $f_j(t|z)$ can be different depending on whether the reduced-form or structural approach is used to form the model.

For the structural approach described above, we derive the implied $f^{s}_j$ using (\ref{model0}):
\begin{eqnarray}
f^{s}_j(t|z; \theta, \beta_1, \beta_2) &=&  -\frac{\partial C\big(S_1(t_1|z;\beta_1), S_2(t_2|z;\beta_2); \theta\big) }{\partial t_1} \bigg|_{t_1=t_2=t} \non\\
&=& - C'_j \big(S_1(t|z;\beta_1), S_2(t|z;\beta_2); \theta\big) S'_j(t|z;\beta_j). \label{fj}
\end{eqnarray}
\nin $C'_j$ is the derivative of $C$ with respect to its $j$-th argument, and $S_j'(t)$ is the derivative of $S_j(t)$ with respect to $t$. Equation (\ref{fj}) shows that $f^{s}_j$ is a complicated mixture of all components in the model, including $C$, $S_1$, and $S_2$. Hence,  $f^{s}_j$ involves all structural parameters ($\theta,\beta_1, \beta_2$). Noteworthy is that $f^{s}_1$ does not solely depend on $\beta_1$, the parameters for the marginal hazard of risk 1, because the marginal survivals of both risks ($S_1, S_2$) and the dependency between them ($C$) jointly shape the incidence rate of risk 1 at any $t$.

We derive the likelihood function for the structural model by substituting $f_j^s$ as given in (\ref{fj}) into the generic likelihood (\ref{like1}):
\begin{eqnarray}
L^{s}_i(\theta, \beta_1, \beta_2 |t_i,z_i)& = & \big[- C'_1 \big(S_1(t|z;\beta_1), S_2(t|z;\beta_2); \theta\big) S'_1(t|z;\beta_1) \big]^{2-\delta_i}
\non \\
&& \hspace{1cm} \times \big[- C'_2 \big(S_1(t|z;\beta_1), S_2(t|z;\beta_2); \theta\big) S'_2(t|z;\beta_2)\big]^{\delta_i -1}. \label{likestr}
\end{eqnarray}
\nin The well-known non-identifiability of the competing risks model (Cox, 1962) prevents identification of the structural parameters. Therefore, the likelihood (\ref{likestr}) cannot be used unless further restrictions are being made on the model.

One popular approach to bypass the non-identifiability is to choose a model for $f_j$ by modelling the CSHs or SDHs. We focus in the following on the CSH
$$h_j(t|z;\alpha_j)  = \text{lim}_{dt\rightarrow 0} \Pr(t\leq T< t+dt,\delta=j| T> t, z; \alpha_j)$$ for $j=1,2$ with parameters $(\alpha_1, \alpha_2)$.  The SDHs are considered in Subsection \ref{sec:SDH}. These alternative approaches can be considered as reduced-form approaches because they do not model the marginal distribution of $T_j$ as in the structural approach. The sub-density functions for $j=1,2$ implied by the CSHs are
\begin{eqnarray}
f^{c}_j(t|z; \alpha_1, \alpha_2) = h_j(t|z; \alpha_j)\exp\big(-H_1(t|z; \alpha_1)- H_2(t|z; \alpha_2)\big), \label{fcsh}
\end{eqnarray}
\nin where $H_j(t|z;\alpha_j) = \int_0^{t} h_j(u|z;\alpha_j) du$ is the cumulative CSH for risk $j=1,2$. The generic likelihood (\ref{like1}) forms also the basis for the approach on the grounds of the CSHs by substituting (\ref{fcsh}) into (\ref{like1}):
\begin{eqnarray}
L^{c}_i(\alpha_1, \alpha_2|t_i,z_i) = \big[h_1(t_i|z_i; \alpha_1)\big]^{2-\delta_i} \big[h_2(t_i|z_i;\alpha_2)\big]^{\delta_i-1}e^{-H_1(t|z; \alpha_1)-H_2(t|z; \alpha_2)}. \label{likecsh}
\end{eqnarray}
\nin $(\alpha_1, \alpha_2)$ are identifiable and can be estimated with this approach.

By comparing $f_1$ in (\ref{fj}) and (\ref{fcsh}), it is apparent that the two models contain different sets of parameters: the structural parameters ($\theta,\beta_1, \beta_2$) in (\ref{fj}) and the reduced-form parameters $(\alpha_1, \alpha_2)$ in (\ref{fcsh}). Analytical one-to-one relationships between these two sets of parameters are generally not available (see, e.g. Emura et al., 2020). It renders it impossible in empirical work to uncover the structural parameters using the identified reduced-form parameters based on the likelihood function in (\ref{likecsh}). This has an important implication. The reduced-form parameters only describe how the observed outcomes $(T,\delta|Z)$ react upon $Z$. They are generally not informative about the effect of $Z$ on the distribution of ($T_1, T_2$). For instance, a rise in the wage level usually reduces the duration of a high school leaver to find a job ($T_1$), as the opportunity cost for job search is higher. At the same time, a rise in the wage level usually increases the duration to enroll in a college ($T_2$), as the opportunity cost for studying is higher. Combining these two effects leads to a reduction in $T$ and a higher incidence of $\delta =1$ (taking a job). The identified $(\alpha_1, \alpha_2)$ can only capture these final outcomes in $(T, \delta)$, but they do not inform us how the wage increase affects $T_1$ and $T_2$ marginally. As a matter of fact, the same set of $(\alpha_1, \alpha_2)$ can be generated by different sets of structural parameters ($\theta,\beta_1, \beta_2$) due to the non-identifiability of the competing risks model. For instance, a reduction in $T$ and a higher incidence of risk 1 can also happen in a case when the duration to enroll in a college decreases (not increases), provided that this reduction is smaller in size than the drop in the duration of job finding. The CSH approach is therefore not informative about the causal effect of wage on college enrollment, although this should be the primary interest of empirical research. In this regard, the CSH approach is a reduced-form model that conveys no causal interpretation, which leaves the underlying mechanisms in a black box. Even more so, we show below that different sets of reduced-form parameters can arise from the same set of structural parameters, when merely the covariates $Z$ have different distributions. Suppose there are two sets, which are generated by the same structural model but with different distributions of the covariates. For example one dataset excludes young individuals as the only difference. The estimated reduced-form parameters can be different in these two samples, simply because the covariates have different marginal distributions. It illustrates the arbitrariness of the reduced-form parameters, making the interpretation of the CSH model even more difficult. This also renders counterfactual analysis on the grounds of these models unviable.

\subsection{Proportionality in CSH modelling \label{sec:CSH}}
In the last section, we have summarised the well-known fact that the CSH model alone is not informative about the data generating process. In this section, we present a new result that the reduced-form approach imposes substantial restrictions on the structural model which can be difficult to justify in an application. Namely, the Cox PH model, which is commonly applied in the context of CSH analysis, is only compatible with the data generating process of the structural model when the copula and marginal hazards possess certain properties. The empirical researcher who adopts a PH CSH model must be therefore aware that this model strongly restricts the DGP. This is shown in the following.

In the PH CSH model, $h_j(t|z)$ is separable in $t$ and $z$:
\begin{eqnarray}
h_j(t|z) = h_{0j}(t)\psi_j(z). \label{coxh}
\end{eqnarray}
\nin An important implication of (\ref{coxh}) is that the hazard ratio evaluated at $Z=z_1$ and $Z=z_2$
\begin{eqnarray}
\frac{h_j(t|z_1)}{h_j(t|z_2)} = \frac{\psi_j(z_1)}{\psi_j(z_2)}. \label{hratio}
\end{eqnarray}
is independent of $t$. It is remarked that $h_j(t|z)$ is a model that the researcher fits to the data. The reduced-form parameters are $\alpha_j=(\alpha_{0j},\alpha_{1j})$, where $\alpha_{0j}$ are the parameters of $h_{0j}(t)$ and $\alpha_{1j}$ enter $\psi_j(z)$.

In the following, we derive $h^{s}_j(t|z)$ the CSH function that is implied by the structural model in (\ref{model0}). For $h^{s}_j(t|z)$ to be compatible with the assumed proportionality of $h_j(t|z;\alpha_j)$ in (\ref{hratio}), it requires substantial restrictions on $C$, $S_1$ and $S_2$. We use the Gumbel copula as illustration.

The Gumbel copula is $C(u,v|\theta) = \exp[-((- \log u)^{\theta} + (-\log v)^{\theta})^{1/\theta}]$ with one dependence parameter $\theta \in [1,\infty]$. The two risks are independent if $\theta =1$, and their correlation increases with $\theta$. Using the Gumbel copula, model (\ref{model0}) becomes
\begin{eqnarray}
\big[-\log S(t|z)\big]^{\theta} = \big[-\log S_1(t|z)\big]^{\theta} + \big[-\log S_2(t|z)\big]^{\theta}. \label{overallS}
\end{eqnarray}
\nin $f^{s}_j$ in (\ref{fj}) becomes $f^{s}_j(t|z) = -S(t|z) \big[-\log S(t|z)\big]^{1-\theta} \big[-\log S_j(t|z)\big]^{\theta-1} \big[S_j(t|z)\big]^{-1} S'_j(t|z)$. The implied CSH $h^{s}_j(t|z)  = f^{s}_j(t|z)/S(t|z)$ is
\begin{eqnarray}
h^{s}_j(t|z) = -\big[-\log S(t|z)\big]^{1-\theta} \big[-\log S_j(t|z)\big]^{\theta-1} \big[S_j(t|z)\big]^{-1}S'_j(t|z). \label{hstr}
\end{eqnarray}
To ensure that $h^{s}_j(t|z)$ in (\ref{hstr}) satisfies the PH assumption in (\ref{coxh}), $S_j(t|z)$ needs to be separable in $t$ and $z$ for $j=1,2$. It is straightforward that this follows from the PH assumption on the marginal hazards:
\begin{eqnarray}
\lambda_j(t|z)=\lambda_{0j}(t)\phi_j(z) \label{lamj}
\end{eqnarray}
for $j=1,2$. It is remarked that $\lambda_j$ is parametrised by the structural parameters, $\beta_j=(\beta_{0j},\beta_{1j})$, where $\beta_{0j}$ are the parameters of $\lambda_{0j}(t)$ and $\beta_{1j}$ enter $\phi_j(z)$. As explained in the previous subsection, the structural parameters $\beta_j$ have a different interpretation than the reduced-form parameters $\alpha_j$. The PH assumption in (\ref{lamj}) simplifies (\ref{hstr}) to
\begin{eqnarray}
h^{s}_j(t|z) = \big[-\log S(t|z)\big]^{1-\theta} \big[\Lambda_{0j}(t)\big]^{\theta-1}\lambda_{0j}(t) \big[\phi_j(z)\big]^{\theta}.\label{hstr2}
\end{eqnarray}
\nin The expression in (\ref{hstr2}) shows that all functions related to $S_j(t|z)$ are separable in $t$ and $z$. It is not enough, though, as $S(t|z)$ must be separable in $t$ and $z$ as well. It will be shown in the following that another restriction is required. In particular, the CSHs for risks 1 and 2 are proportional to each other. This is what is called risk-proportionality (RP) in the literature (Kalbfleisch and Prentice, 2002; Lo et al., 2024). To see this, the overall survival function is written as $-\log S(t|z)=H(t|z)$ and (\ref{coxh}) is used, such that
\begin{eqnarray}
H(t|z) = H_1(t|z) + H_2(t|z) =  H_{01}(t)\psi_1(z) + H_{02}(t)\psi_2(z), \label{Htz}
\end{eqnarray}
\nin where $H_{0j}(t) =\int_0^{t} h_{0j}(u) du$ is the cumulative baseline CSH for risk $j=1,2$.

$H_{01}(t)$ and $H_{02}(t)$ in (\ref{Htz}) are allowed to have different functional forms in the commonly used PH CSHs models. But the separability of $H(t|z)$ in $t$ and $z$ in (\ref{Htz}) requires the RP restriction, such that:
\begin{eqnarray}
H_{02}(t) = e^{\gamma}H_{01}(t) \hspace{0.5cm} \text{ or  }  \hspace{0.5cm} h_{02}(t) = e^{\gamma}h_{01}(t), \label{prisk}
\end{eqnarray}
\nin for any real number $\gamma$. Put (\ref{prisk}) into (\ref{Htz}), we obtain
\begin{eqnarray}
H(t|z) = H_{01}(t)\big[\psi_1(z) +\exp(\gamma)\psi_2(z)\big] = H_{0}(t) \psi^*(z), \label{Htz2}
\end{eqnarray}
\nin where we define $H_0(t) = H_{01}(t)$ and $\psi^*(z)=\psi_1(z) +\exp(\gamma)\psi_2(z)$  for convenience. We can now further simplify $h^{s}_j(t|z)$ in (\ref{hstr2}) by using (\ref{Htz2}). This step makes $t$ and $z$ separable in $h^{s}_j(t|z)$, as
\begin{eqnarray}
h^{s}_j(t|z) = \bigg(\lambda_{0j}(t)\big[\Lambda_{0j}(t)/H_0(t)\big]^{\theta-1}\bigg) \times \bigg(\big[\psi^*(z)\big]^{1-\theta}  \big[\phi_j(z)\big]^{\theta}\bigg).\label{hstr3}
\end{eqnarray}
\nin The hazard ratio for $h^{s}_j(t|z)$ evaluated at $Z=z_1$ and $Z=z_2$ is then
\begin{eqnarray}
\frac{h^{s}_j(t|z_1)}{h^{s}_j(t|z_2)} = \bigg[\frac{\psi^*(z_1)}{\psi^*(z_2)}\bigg]^{1-\theta}\bigg[\frac{\phi_j(z_1)}{\phi_j(z_2)}\bigg]^{\theta} \label{hratio2}
\end{eqnarray}
\nin which does not depend on $t$ and therefore $h_j^s(t|z)$ satisfies the PH restriction in analogy to the assumed proportional hazards structure for the CSH model in (\ref{coxh}). The following theorem summarises the previous observations.

\theorem{The CSH $h^{s}_j(t|z)$ in (\ref{hstr}), derived from the structural model, satisfies the PH property in (\ref{coxh}) for all $t$ and $Z$ if\\
(i)   $h_j(t|z)$ satisfies the PH property in (\ref{coxh});\\
(ii)  $h_{0j}(t|z)$ satisfies the RP property in (\ref{prisk});\\
(iii)  $\lambda_j(t|z)$ satisfies the PH property in (\ref{lamj}); and \\
(iv)  $C$ is a Gumbel copula.\label{ass1}}

Under this set of restrictions, the assumed PH CSH model in (\ref{coxh}) is compatible with a structural competing risks model in (\ref{model0}). It is worth remarking on that the PH property for the CSH in (i) alone is insufficient. Properties (i) - (iv) in Theorem \ref{ass1} are sufficient but not necessary conditions, which warrant some discussion. Condition (iv) implies that the Gumbel copula is an appropriate choice when working with a PH CSHs model. We have tried to derive an equivalent result for other well-known copulas such as Clayton or Frank, but did not succeed. It is shown in Appendix II that the PH CSHs model is generally incompatible with a Clayton copula except of the trivial case of independent risks.  Condition (ii) of Theorem \ref{ass1} shows that $h_1(t|x)$ and $h_2(t|x)$ cannot be arbitrarily specified as they must comply with the RP property. Condition (ii) leads to another interesting implication that we consider next.

Due to the RP property (ii), we can wrtie the ratio of the CSHs for risks 1 and 2 using $(\ref{coxh})$ as
\begin{eqnarray}
\frac{h_2(t|z)}{h_1(t|z)} =\frac{h_{02}(t)}{h_{01}(t)} \frac{\psi_2(z)}{\psi_1(z)} = e^{\gamma}\frac{\psi_2(z)}{\psi_1(z)} .\label{hratio3}
\end{eqnarray}
\nin This is independent of $t$ due to the RP property in (\ref{prisk}). For this CSH model to be compatible with the structural model,  the same proportionality must hold for $h^{s}_j(t|z)$ as well. Dividing  $h^{s}_2(t|z)$ by  $h^{s}_1(t|z)$ in (\ref{hstr3}) gives
\begin{eqnarray}
\frac{h^{s}_2(t|z)}{h^{s}_1(t|z)} = \bigg[\frac{\lambda_{02}(t)}{\lambda_{01}(t)}\bigg] \bigg[ \frac{\Lambda_{02}(t)}{\Lambda_{01}(t)} \bigg]^{\theta-1} \times \bigg[ \frac{\phi_2(z)}{\phi_1(z)}\bigg]^{\theta}. \label{hratio32}
\end{eqnarray}
\nin To eliminate $t$ in (\ref{hratio32}), it requires a similar risk-proportionality restriction for the marginal hazard functions. Namely, for any real number $\varsigma$,
\begin{eqnarray}
\Lambda_{02}(t) = e^{\varsigma}\Lambda_{01}(t)  \hspace{0.5cm} \text{ or  }  \hspace{0.5cm} \lambda_{02}(t) = e^{\varsigma}\lambda_{01}(t). \label{prisk2}
\end{eqnarray}
Putting (\ref{prisk2}) into (\ref{hratio32}) gives
\begin{eqnarray}
\frac{h^{s}_2(t|z)}{h^{s}_1(t|z)} =  e^{\varsigma\theta} \bigg[\frac{\phi_2(z)}{\phi_1(z)}\bigg]^{\theta},\label{hratio4}
\end{eqnarray}
\nin which no longer depends on $t$ and the risk-proportionality holds for $h_1^s$ and $h_2^s$.

\corollary{Under the restrictions (i)-(iv) of Theorem \ref{ass1}, $\lambda_{0j}(t|z)$ must satisfy the RP property in (\ref{prisk2}). \label{cor1}}

This corollary suggests that RP of on the CSHs implies the same RP property the the marginal hazards. The analytical relationship between the CSHs and the marginal hazards can be further specified under Assumption \ref{ass1} and Corollary \ref{cor1}. Equating (\ref{hratio3}) and (\ref{hratio4}) gives
\begin{eqnarray}
e^{\gamma}\frac{\psi_2(z)}{\psi_1(z)} =  e^{\varsigma\theta}\bigg[ \frac{\phi_2(z)}{\phi_1(z)}\bigg]^{\theta}.\label{hratio6}
\end{eqnarray}
\nin For (\ref{hratio6}) to hold for any $z$ and $\theta$, it must be that
\begin{eqnarray}
 \gamma &=& \varsigma \theta,\label{res1} \\
 \psi_2(z)/\psi_1(z) &=&  [\phi_2(z)/\phi_1(z)]^{\theta}.\label{res2}
\end{eqnarray}
\nin (\ref{res1}) shows that there is a direct relationship between the risk proportionality of CSHs $(\gamma)$ and the marginal hazards ($\varsigma)$. This relationship is governed by $\theta$. In the case of independence ($\theta=1$), they are the same.

Next, by equating the CSH functions in (\ref{coxh}) and (\ref{hstr3}) gives
\begin{eqnarray}
h_{0j}(t) &=&  \lambda_{0j}(t) [\Lambda_{0j}(t)/H_0(t)]^{\theta -1}, \label{res0} \\
\psi_{j}(z) &=& [\psi^*(z)]^{1-\theta}[\phi_j(z)]^{\theta} . \label{res3}
\end{eqnarray}
\nin Rearranging (\ref{res0}) and integrating both sides with respect to $t$, we obtain
\begin{eqnarray}
H_{01}(t) =  \Lambda_{01}(t). \label{res5}
\end{eqnarray}
\nin The baseline cumulative CSH function and the baseline cumulative marginal hazard function are identical for risk 1.
For risk 2, the two functions have a fixed ratio:
\begin{eqnarray}
H_{02}(t) =  \exp[\gamma(1- \theta^{-1})] \Lambda_{02}(t). \label{res4}
\end{eqnarray}
\nin Equations (\ref{res5}) and (\ref{res4}) imply that the CSHs and the marginal hazards have the same baseline hazard except for a scaling constant.

It is also possible to establish the link between the covariate function of the CSH $\psi_{j}(z)$ and the covariate function of the marginal hazards $\phi_{j}(z)$ for $j=1,2$. From (\ref{res2}), $\psi_2 (z) = \psi_1(z) (\phi_2(z)/\phi_1(z))^{\theta}$. Substitute this into (\ref{res3}) and use $\psi^*(z)=\psi_1(z) +\exp(\gamma)\psi_2(z)$ to obtain
\begin{eqnarray}
\psi_{1}(z) = \phi^{\theta}_{1}(z) [\phi^{\theta}_1(z) +\exp(\gamma)\phi^{\theta}_2(z)]^{1/\theta-1} \label{res6}
\end{eqnarray}
\nin for risk 1. For risk 2 one can show
\begin{eqnarray}
\psi_{2}(z) = \phi^{\theta}_{2}(z) [\phi^{\theta}_1(z) +\exp(\gamma)\phi^{\theta}_2(z)]^{1/\theta-1}.\label{res62}
\end{eqnarray}
\nin It is remarked that $\psi_j$ is a function with parameters $\alpha_{1j}$, while $\phi_j$ is a function with  parameters $\beta_j$. From (\ref{res6}), $\psi_1(z)$ is a complicated function involving $\phi_1(z)$, $\phi_2(z)$ as well as the parameters $\gamma$ and $\theta$. It confirms that $\alpha_{1j}$ of $\psi_1$ is shaped by all structural parameters $(\beta_{11},\beta_{12},\gamma,\theta)$. Since $\alpha_{1j}$ does not correspond to $\beta_{1j}$ alone, it cannot be inferred as any causal effect of $Z$ on the distribution of $T_1$.

We summarize the above relationships in a corollary.

\corollary{Under restrictions (i)-(iv) of Theorem \ref{ass1}, $h_j(t|z)$ and $\lambda_j(t|z)$ for $j=1,2$, and for all $t$ and $z$, are related in the following ways:\\
(v) $\gamma = \varsigma \theta$; \\
(vi)  $H_{01}(t) =  \Lambda_{01}(t)$;\\
(vii)  $H_{02}(t) = \exp[\gamma(1-\theta^{-1})] \Lambda_{02}(t)$; and\\
(viii) $\psi_{j}(z) = \phi^{\theta}_{j}(z) [\phi^{\theta}_1(z) +\exp(\gamma)\phi^{\theta}_2(z)]^{\theta^{-1}-1}$.\label{cor2}}

Conditions (v)-(viii) are additional restrictions on $h_{0j}(t)$ and $\lambda_{0j}(t)$ for being compatible with a well defined structural competing risks model.

Another restriction on $\psi_j(z)$ can be derived if $\phi_j(z) = \exp(z\beta_{1j})$. This is the default specification of the covariate function in applied research. It is known that this choice of covariate function implies a constant log hazard ratio (LHR) for the marginal hazard, defined as $\text{LHR} = \partial \log \phi_j(z) / \partial z = \beta_{1j}$. The LHR is constant, as it does not dependent on $Z$. Equations (\ref{res6}) and (\ref{res62}) have established the link between $\psi_j(z)$ and $\phi_j(z)$. When using the above exponential specification for $\phi_j$, the right hand sides of these equations are rewriten as:
\begin{eqnarray}
\psi_{1}(z) &= &\exp(z\beta_{11}) \bigg[1 +  e^{\gamma} \exp(z(\beta_{12}-\beta_{11})\theta) \bigg]^{1/\theta-1},\label{res7} \\
\psi_{2}(z) &= &\exp[z(\beta_{11}(1-\theta) +\beta_{12}\theta)] \bigg[1 +  e^{\gamma} \exp(z(\beta_{12}-\beta_{11})\theta) \bigg]^{1/\theta-1}. \label{res72}
\end{eqnarray}
It is obvious that $\psi_j(z)$ in (\ref{res7}) and (\ref{res72}) cannot be the exponential covariate function $\psi_{j}(z) = \exp(\alpha_{1j} z)$. In this case, the log hazard ratio for $\psi_{j}(z)$ would be as $\text{LHR} = \partial \log \psi_j(z) / \partial z=\alpha_{11}$. Indeed, the log hazard ratio for $\psi_{j}(z)$ implied by the structural models in (\ref{res7}) and (\ref{res72}) is not constant. For (\ref{res7}) it is
\begin{eqnarray}
\frac{\partial \log \psi_{1}(z)}{\partial z} = \beta_{11} - (\theta-1) (\beta_{12} - \beta_{11}) \frac{\exp(\varsigma \theta) \exp(z (\beta_{12} - \beta_{11}) \theta)}{1+  e^{\varsigma \theta} \exp(z (\beta_{12} - \beta_{11}) \theta)}, \label{comp1}
\end{eqnarray}
\nin which is a function of $Z$.

What can be inferred from the resulting $\alpha_{1j}$ when a researcher applies the PH CSH model together with the exponential covariate function? The resulting $\alpha_{1j}$ can be viewed as a sample average of the LHR. In the case of $j=1$ it is given by
\begin{eqnarray}
\alpha_{11} &:=& \int \frac{\partial \log \psi_{1}(z)}{\partial z} d F(z) \non\\
&=&  \beta_{11} - (\theta-1) (\beta_{12} - \beta_{11})\exp(\varsigma \theta) \int  \frac{ \exp(z (\beta_{12} - \beta_{11}) \theta)}{1+  e^{\varsigma \theta} \exp(z (\beta_{12} - \beta_{11}) \theta)} dF(z), \label{alpha11}
\end{eqnarray}
\nin which depends not just the structural parameters but also on $F(z)$, the marginal distributions of $Z$. Suppose there are two populations with the same structural model but with different marginal distributions of $Z$. The resulting $\alpha_{1j}$ for the two populations are different because $Z$ has different distributions. It renders not only interpretability of $\alpha_{1j}$ difficult, but also affects out-of-sample predictions based on $\alpha_{1j}$. This is because the value of $\alpha_{1j}$ is contingent upon a specific distribution of $Z$, which may take another value in another population with different $F(z)$. We show with the help of simulations in Section \ref{sec:sim} that this can strongly affect the results.

Is the assumption of a constant LHR for the PH CSHs then generally incompatible with a competing risks model? In the following, we show that it is a special case under certain restrictions. We consider two scenarios. The first is $\gamma = -\infty$. (\ref{res7}) and (\ref{res72}) then become
\begin{eqnarray}
\psi_{1}(z) &: = \exp(z\alpha_{11}) =& \exp(z\beta_{11}), \label{case1} \\
\psi_{2}(z) &: =   \exp(z\alpha_{12}) =& \exp[z(\beta_{11}(1-\theta) + \beta_{12} \theta)], \label{case2}
\end{eqnarray}
\nin with $\alpha_{11} = \beta_{11}$, and $\alpha_{12}=\beta_{11}(1-\theta) + \beta_{12} \theta$. When the covariate functions of the CSHs are exponential functions and by using (\ref{res5}) and (\ref{res4}), we obtain
\begin{eqnarray}
h_{1}(t|z) &=& \lambda_1(t;z), \label{case12} \\
h_{2}(t|z) &=& \lim_{\gamma \to -\infty}\lambda_2(t;z)\exp(\gamma) \exp[z(\beta_{12}-\beta_{11})(\theta-1)] = 0 . \label{case22}
\end{eqnarray}
\nin In this scenario, the competing risks model degenerates into a single risk model as there is never an incidence for risk 2. The second scenario is when $\theta =1$ (independent risks). In this case, (\ref{res7}) and (\ref{res72}) become
\begin{eqnarray}
\psi_{j}(z) = \exp(z\beta_{1j}) \label{res9}
\end{eqnarray}
\nin for $j=1,2$. Let the covariate function of the CSHs be the exponential function again. For this to hold for all $z$, $\psi_{j}(z) = \exp(z\alpha_{1j})$ and therefore $\alpha_{1j}=\beta_{1j}$. Using again (\ref{res5}) and (\ref{res4}), $h_j(t|z) = \lambda_j(t|z)$ for $j=1,2$. The CSHs are exactly the marginal hazards when risks are independent. Both scenarios are special cases that are not interesting in a dependent competing risks setup. To summarize, we have the following corollary.

\corollary{Under Theorem \ref{ass1}, a PH CSH model in (\ref{coxh}) in general cannot have a constant log hazard ratio or, specifically, $\psi_j(z)$ does not have an exponential form.\label{cor3}}


Lastly, we point to the fact that not even the sign of the log hazard ratio of the CSH is informative about the sign of the causal effect of $Z$ on the marginal distribution of $T_j$. Suppose the fitted $\psi_{j}(z) = \exp(z\alpha_{1j})$ and the DGP uses $\phi_j(z) = \exp(z\beta_{1j})$. It is possible that the log hazard ratio for the CSH for risk 1 can be negative ($\alpha_{11}<0$) while the causal effect is positive ($\beta_{11}>0$). Suppose $\beta_{11} > 0$, (\ref{comp1}) can still be negative under the following conditions: (i) $\beta_{12}-\beta_{11}$ is large enough, i.e., the marginal effect of risk 2 ($\beta_{12}$) is more positive than that of risk 1 ($\beta_{11}$). It leads to faster exits to risk 2 than risk 1, crowding out the exits to risk 1, even though the duration for risk 1 is shortened due to a positive $\beta_1$. It results in a negative $\alpha_{11}$. (ii) $\varsigma$ is large enough, i.e., the baseline hazard for risk 2 ($\lambda_{02}$) is much larger than that for risk 1, $(\lambda_{01})$, crowding out again the exits to risk 1. (iii) $\theta$ is more positive, i.e., there is a stronger positive dependence between risks 1 and 2. A shorter $T_1$ is accompanied by a shorter $T_2$ as well, offsetting the effect of a positive $\beta_{11}$. It follows that increasing the value of $Z$ can lead to a reduction in the CSH for risk 1, even when it increases the marginal hazard for risk 1. It is therefore generally misleading to interpret the sign of  $\alpha_{1j}$ as the direction of the causal effect of $Z$ on $T_j$ for $j=1,2$. This is something that should be kept in mind when a researchers works with a fitted Cox CSH model in an application.

\subsection{Proportionality assumption in SDH modelling \label{sec:SDH}}
We repeat the analysis of the previous subsection for a model with an assumed Cox PH SDHs model. The SDH is defined as
$$d_j(t|z) = \text{lim}_{dt\rightarrow 0} \Pr(t\leq T_j< t+dt, \delta=j | \{T> t\} \cup \{T\leq t \cap \delta \neq j\}, z).$$
\nin for $j=1,2$. It is directly linked to the incidence function $f_j(t|z)$ through
\begin{eqnarray}
d_j(t|z) = \frac{f_j(t|z)}{1-\int_0^t f_j(s|z)ds}. \label{dj}
\end{eqnarray}
\nin The Cox PH SDHs model for $j=1,2$ is
\begin{eqnarray}
d_j(t|z) &=& d_{j0}(t)\rho_j(z)  \label{cox_dj}
\end{eqnarray}
with $d_{j0}(t)$ is the baseline SDH and $\rho_j(z)$ is the covariate function. This separability restriction implies once again that the ratio of two $d_j(t|z)$ evaluated at different $z$ is independent of $t$:
\begin{eqnarray}
\frac{d_j(t|z_1)}{d_j(t|z_2)} = \frac{\rho_j(z_1)}{\rho_j(z_2)} \label{coxdj}
\end{eqnarray}
for all $z_1,z_2$ and $j=1,2$. Fine and Gray (1999) adopted $\rho_j = \exp(z_j)$ for $j=1,2$, which gives the Cox model for the SDH functions. In the following, we proceed in the same way as we did for the CSHs. By substituting the $f^s_j(t|z)$ that is implied by the structural model into equation (\ref{dj}) gives
\begin{eqnarray}
d^s_j(t|z) = \frac{f^s_j(t|z)}{1-\int_0^t f^s_j(s|z)ds}. \label{djs}
\end{eqnarray}
We then derive the conditions on $\lambda_j$ and $C$ that make $d_j^s$ compatible with the separability restriction in (\ref{cox_dj}).

Unlike the CSH in (\ref{hstr}), the SDH $d_j(t|z)$ in (\ref{dj}) requires $f^s_j(t|z)$ explicitly, so we derive the latter using the structural model as an intermediate step. Using the Gumbel copula in (\ref{overallS}), the implied $f^{s}_j$ in (\ref{fj}) is again $f^{s}_j(t|z) = -S(t|z) \big[-\log S(t|z)\big]^{1-\theta} \big[-\log S_j(t|z)\big]^{\theta-1} \big[S_j(t|z)\big]^{-1} S'_j(t|z)$. Substituting $S(t|z) = C(S_1(t|z),S_2(t|z))$ and $S_j(t|z) = \exp(-\Lambda_{0j}(t)\phi_j(z))$ into $f^{s}_j(t|z)$, the implied $f^s_j(t|z)$ becomes:
\begin{eqnarray}
f^s_1(t|z) &=& \exp\{-[\Lambda^{\theta}_1(t|z)+ \Lambda^{\theta}_2(t|z)]^{1/\theta}\} \non\\
& & \times [\Lambda^{\theta}_1(t|z)+ \Lambda^{\theta}_2(t|z)]^{1/\theta-1} \times \Lambda^{\theta-1}_1(t|z) \lambda_1(t|z).  \label{fstrd01}
\end{eqnarray}
\nin The same RP property in (\ref{prisk2}) is required to extract the common factor for $\Lambda_1(t|z)$ and $\Lambda_2(t|z)$ inside the power function. By doing so, the implied $f^s_j(t|z)$  can be simplified as:
\begin{eqnarray}
f^s_1(t|z) = \lambda_{1}(t|z)(1+\eta(\varsigma, z)^{\theta})^{1/\theta-1}\exp[-\Lambda_1(t|z)(1+\eta(\varsigma,z)^{\theta})^{1/\theta}],  \label{fstrd1}
\end{eqnarray}
\nin where $\eta(\varsigma,z)$ is obtained using the RP of $\Lambda_{0j}(t)$ in (\ref{prisk2}), such that
\begin{eqnarray}
\Lambda_2(t|z) = \frac{\phi_2(z)}{\phi_1(z)}e^{\varsigma} \Lambda_1(t|z) = \eta(\varsigma,z)  \Lambda_1(t|z).
\end{eqnarray}
\nin Let $\tilde{\phi}_{1,\theta}(z) = (1+\eta(\varsigma, z)^{\theta})^{1/\theta}$ be the new covariate function. (\ref{fstrd1}) then becomes
\begin{eqnarray}
f^s_1(t|z) = \lambda_{1}(t|z)\tilde{\phi}^{1-\theta}_{1,\theta}(z)\exp[-\Lambda_1(t|z)\tilde{\phi}_{1,\theta}(z)].  \label{fstrd02}
\end{eqnarray}
\nin It is easy to see that $\int_0^t f_1(s|z)ds = \tilde{\phi}^{-\theta}_{1,\theta}(z)[1-\exp(-\Lambda_1(t|z)\tilde{\phi}_1(z))]$. $d^s_1(t|z)$ in (\ref{djs}) then becomes
\begin{eqnarray}
d^s_1(t|z) &=&  \frac{\lambda_{1}(t|z)\tilde{\phi}_{1,\theta}(z)}{1+\tilde{\phi}^{\theta}_{1,\theta}(z)[1- \exp(-\Lambda_1(t|z)\tilde{\phi}_{1,\theta}(z))]} .
\end{eqnarray}
\nin We use the PH assumption for $\lambda_j(t|z)$ in (\ref{lamj}) to separate $t$ and $z$ in $d^s_1(t|z)$. By doing so, the ratio $d^s_1(t|z_1)/d^s_1(t|z_2)$ simplifies to
\begin{eqnarray}
\frac{d^s_1(t|z_1)}{d^s_1(t|z_2)} &=&  \frac{\phi_1(z_1)}{\phi_1(z_2)} \times \frac{\tilde{\phi}_{1,\theta}(z_1)}{\tilde{\phi}_{1,\theta}(z_2)} \frac{1+\tilde{\phi}^{\theta}_{1,\theta}(z_2)[1-\exp(-\Lambda_{01}(t|z_2)\phi_1(z_2)\tilde{\phi}_{1,\theta}(z_2))]}{1+\tilde{\phi}^{\theta}_{1,\theta}(z_1)[1-\exp(-\Lambda_{01}(t|z_1)\phi_1(z_1)\tilde{\phi}_{1,\theta}(z_1)])}. \label{dratio}
\end{eqnarray}
\nin In contrast to the case of the CSHs, this ratio still depends on $t$ due to the presence of $\Lambda_{01}(t)$. There is only one special case that guarantees the validity of the PH property for the SDHs model, which requires that $t$ disappears from the right hand side of (\ref{dratio}). Namely, $\theta\rightarrow \infty$. By L'Hoptial's rule, $d^s_1(t|z_1)/ d^s_1(t|z_2) \to \phi_1(z_1)/\phi_2(z_2)$ and the PH property is satisfied. In other words, the SDHs can only have the PH property as in (\ref{cox_dj}), when the two risks have perfect positive dependence. This corresponds to that the copula attains its Fr\'{e}chet-Hoeffding upper bound, i.e., $C(u,v) = \min(u,v)$.


\theorem{The SDH $d^s_j(t|z)$ in (\ref{djs}), derived from the structural model, satisfies the PH assumption in (\ref{cox_dj}), if\\
(ix) $\lambda_j(t|z)$ satisfies the PH property in (\ref{lamj}); \\
(x) $\lambda_j(t|z)$ satisfies the RP property in (\ref{prisk2}); and  \\
(xi) $C$ is a Fr\'{e}chet-Hoeffding upper bound copula.\label{cor5}}

As for Theorem \ref{ass1}, Theorem \ref{cor5} presents sufficient conditions. Another sufficient conditions for the PH assumption is referred to Example 1 of Emura et al. (2020) that also requires conditions (ix) and (xi). In either case, the use of a PH SDHs model has severe implications for the DGP and these restrictions are stronger than in the case of the CSHs models. We therefore focus on the latter in the remainder of this paper.

\section{Partial likelihood estimation \label{sec:est}}
For the estimation of proportional hazards models, it is convenient to craft partial likelihoods. Kalbfleish and Prentice (2002) consider estimation under risk proportionality restrictions and provide a partial likelihood approach where the reduced-form parameters $(\alpha_{11}, \alpha_{12}, \gamma)$ of the PH CSHs model with $\psi_j(z) = \exp(z\alpha_{1j})$ and $h_{02}(t) = e^{\gamma} h_{01}(t)$ are estimated. As shown in Section \ref{sec:CSH}, the exponential specification of the covariate function $\psi_j(z)$ implies a constant log hazard ratio, which is incompatible with the structural model unless the competing risks are independent. A solution is to avoid estimation of the reduced-form PH CSHs model, but to focus on the structural model instead. In this section, we show that the structural parameters $(\theta, \varsigma, \beta_{11}, \beta_{12})$ can be estimated by the method of partial likelihood as well, provided the conditions of Theorem \ref{ass1} hold. We can retain the flexibility of the partial likelihood method in the sense that the marginal baseline hazard $\lambda_{0j}(t;\beta_{0j})$ can be left unspecified and is nonparametrically estimated in a second step. We can, at the same time, apply the standard exponential specification to the covariate function of the marginal hazards.

The partial likelihood under the RP property for the CSHs has been considered by Kalbfleish and Prentice (2002) and Lo et al. (2024):
\begin{eqnarray}
L^{p}(\alpha; t,\delta,z) = \prod_{i=1}^n   \frac{\sum_{j=1}^2 \1(\delta_i=j)h_{j}(t_i|z_i; \alpha_j)}{ \sum_{s\in R(t_i)} h_1(t_s|z_s; \alpha_1)+h_2(t_s|z_s; \alpha_2)}, \label{PL0}
\end{eqnarray}
\nin with $\alpha=(\alpha_1,\alpha_2)$ and $\1$ is the indicator function that is 1 if the condition is met and 0 otherwise. The contribution of observation $i$ is the conditional probability that $i$ fails of any risk ($\delta_i$) at $t_i$ given the risk set at $t_i$ ($R(t_i)$), which are all observations which have not failed for either risk. Note that $h_1(t|z) = h_{01}(t)\psi_1(z)$ and $h_2(t|z) = h_{02}(t)\psi_2(z) = e^{\gamma}\psi_2(z)$ under (\ref{prisk}). We can specify $h_j(t|z) = h_{01}(t)e^{\gamma_j}\psi_j(z)$, where $\gamma_1 =0$ and $\gamma_2 = \gamma$. By substituting this model for $h_j(t|z)$ into (\ref{PL0}), $h_{01}(t)$ is eliminated and the resulting partial likelihood no longer depends on $t$:
\begin{eqnarray}
L^{p}(\alpha;\delta,z) = \prod_{i=1}^n \frac{\sum_{j=1}^2 \1(\delta_i=j)\left[e^{\gamma_j}\psi_{j}(z_i;\alpha_{1j})\right]}{\sum_{s\in R(t_i)} e^{\gamma_1}\psi_1(z_s;\alpha_{11}) +e^{\gamma_2}\psi_2(z_s;\alpha_{12})}, \label{PL2}
\end{eqnarray}
\nin with $\alpha_j=(\gamma_j,\alpha_{1j})$ for $j=1,2$. Since $\psi_j(z)$ cannot have the usual exponential form in the CSH model,
the next step is to find a model for $\psi_j(z)$ that is implied by the structural models as defined in (\ref{res7}) and (\ref{res72}). By doing so, the likelihood for the reduced-form parameters $(\gamma, \alpha_{11}, \alpha_{12})$  is reparameterised into a likelihood for the structural model with parameters $\Sigma=(\theta, \gamma, \beta_{11}, \beta_{12})$. To further simplify the notation, we define
\begin{eqnarray}
\beta^*_{11}  &=& \beta_{11} , \label{b1s} \\
\beta^*_{12}(\theta, \gamma,  \beta_{11}, \beta_{12}) &=& \beta_{11}(1-\theta) + \beta_{12}\theta, \label{b2s} \\
A(z;\theta, \gamma, \beta_{11}, \beta_{12}) &=& [1+\exp(\gamma)\exp(z(\beta_{12}-\beta_{11})\theta)]^{1/\theta -1}. \label{Az}
\end{eqnarray}
\nin The likelihood in (\ref{PL2}) becomes
\begin{eqnarray}
&& L^{p}(\Sigma; t,\delta,z) \non\\
&=& \prod_{i=1}^n   \frac{\exp\left[\sum_{j=1}^2 \1(\delta_i=j)\left(\gamma_{j}+z_i\beta^*_{1j}\right) + \log A(z_i)\right]}{ \sum_{s\in R(t_i)} \exp[\gamma_{1}+z_s\beta^*_{11} + \log A(z_s)] + \exp[\gamma_{2}+z_s\beta^*_{12} + \log A(z_s)]}. \label{PL3}
\end{eqnarray}
\nin All structural parameters, except the nuisance parameters in the baseline hazards $(\beta_{01},\beta_{02})$, can be estimated by partial likelihood using (\ref{PL3}). The log likelihood of (\ref{PL3}) is
\begin{eqnarray}
\log L^{p}(\Sigma; t,\delta,z) & = & \sum_{i=1}^n \sum_{j=1}^2 \1(\delta_i=j)\left( \gamma_{j}+z_i\beta^*_{1\j}\right) + \log A(z_i) \non\\
&&  - \log \bigg[\sum_{s\in R(t_i)} \sum_{k=1}^2 \exp[\gamma_{k}+z_s\beta^*_{1k} + \log A(z_s)] \bigg]. \label{logPL2}
\end{eqnarray}
As outlined in Kalbfleish and Prentice (2002), log partial likelihoods under the RP property can be estimated with a restructured dataset, which transforms the two-risk model into a single risk. We consider a simple example for illustrations. The original dataset has two observations, $i=1,2$. The first observation with $z_1$ fails for risk 1 at $T_1$ and the second observation with $z_2$ fails for risk 2 at $T_2$. The original dataset is
\[
 \begin{blockarray}{ cccc }
 i & \delta_i & T_i & z_i   \\
\begin{block}{ [cccc] }
1 & 1 & T_1 & z_1\\
2 & 2 & T_2 & z_2  \\
\end{block}
 \end{blockarray}.
\]
\nin We duplicate each observation so that the size of the sample becomes $2n$.

 \[
 \begin{blockarray}{ cccccccccc }
 i & \tilde{i} & J & \delta_i & \tilde{\delta}_i & \tilde{T}_i & \tilde{z}_{1i} & \tilde{z}_{2i} & \tilde{z}_{3i} & \tilde{z}_{4i}  \\
\begin{block}{ [cccccccccc] }
1 & 1 & 1 & 1 & 1 & T_1 & 1 & z_1 & 0 & 0\\
1 & 2 & 2 & 1 & 0 & T_1 & 0 & 0   & 1 & z_1\\
2 & 3 & 1 & 2 & 0 & T_2 & 1 & z_2 & 0 & 0 \\
2 & 4 & 2 & 2 & 1 & T_2 & 0 & 0   & 1 & z_2 \\
\end{block}
 \end{blockarray}
\]
The procedure also includes the creation of several new variables. The risk index $J$ in the first row for each $i$ is one and in the second row is two. $\tilde{\delta}_i$ is the new failure indicator in this single-risk model. It takes the value 1 when there is a failure and zero in the case of independent right-censoring. $\tilde{\delta}_i$ = 1 whenever $\delta_i = J$, e.g., in rows 1 and 4. $\tilde{\delta}_i$ = 0 whenever $\delta_i  \neq J$, e.g., in rows 2 and 3. The new covariate vector is $\tilde{Z} =  (\tilde{Z}_{1i}, \tilde{Z}_{2i},  \tilde{Z}_{3i}, \tilde{Z}_{4i})$. We define $\tilde{Z} =  (1, z_i, 0,0)$ in the first row and $\tilde{Z} =  (0, 0, 1, z_i)$ in the second row for each duplicated observation $i$. The corresponding vector of parameters is $\omega = (\gamma_1, \beta^*_{11}, \gamma_2, \beta^*_{12})$.

\nin We can estimate $\omega$ and the parameters in $A(z)$ from this restructured dataset using the following log likelihood function under the constraint $\gamma_1 =0$:
 \begin{eqnarray}
\log L^{p2}(\Sigma; \tilde{t},\tilde{\delta},\tilde{z}) & = & \sum_{\tilde{i}:\tilde{\delta}_i =1 }  \tilde{z}_{\tilde{i}}\omega + \log A(\tilde{z}_{2\tilde{i}}+\tilde{z}_{4\tilde{i}}) \non\\
&&   - \log \bigg[\sum_{s\in R(\tilde{t}_{\tilde{i}})} \sum_{k=1}^2 \exp[ \tilde{z}_s\omega + \log A(\tilde{z}_{2s}+\tilde{z}_{4s})] \bigg]. \label{logPL3}
\end{eqnarray}
\nin It is remarked that the structural parameters $(\theta, \gamma,   \beta_{11}, \beta_{12})$ are linked to  $\omega$ and  $A(z)$ through (\ref{b1s}), (\ref{b2s}) and (\ref{Az}). By reformulating the likelihood (\ref{logPL3}), one can directly estimate $(\theta, \gamma, \beta_{11}, \beta_{12})$. 
The asymptotic properties of this partial likelihood estimation approach can be found in Kalbfleish and Prentice (2002, section 5.7). The estimator for $(\gamma, \beta)$ is $\sqrt{n}-$consistent.

\section{Simulations \label{sec:sim}}
The competing risks model with structural parameters $(\theta, \gamma, \beta_{01}, \beta_{02}, \beta_{11}, \beta_{22})$ is simulated such that it satisfies the restrictions of Theorem \ref{ass1}. There is one continuous covariate $Z$ with $Z\sim N(0,2)$. The marginal hazard functions have an exponential baseline, and their covariate functions have the exponential form, i.e., $\lambda_1(t|z) = \beta_{01}\exp(z\beta_{11})$ and $\lambda_2(t|z) = \beta_{02}\exp(z\beta_{12})$. We set $\beta_{02} = e^{\varsigma}\beta_{01}$ because of Corollary \ref{cor1}, and $\gamma = \varsigma \theta$ because of Corollary \ref{cor2}, $\beta_{02}$ is fixed by the chosen $\beta_{01}$, $\gamma$\, and $\theta$. We set $\gamma = 0.5$, $\beta_{01}$ = 1, $\beta_{11}$ = 1, and $\beta_{12}$ = 2. The resulting $\delta$ has a mean value of 1.58, meaning that 42\% of the observations fail with risk 1 and 58\% fail with risk 2. To assess how the method works for different degrees of dependence, three  values of $\theta$ are considered, namely 1.11, 2, and 10. The corresponding Kendall's $\tau$ $(=1-1/\theta)$ are 0.1, 0.5, and 0.9, respectively. They cover the full range of possible rank correlations from slightly positively correlated risks to strongly positively correlated risks. We simulate 500 samples with independent observations for three sample sizes $n=100, 200, 400$. These samples sizes are rather small for competing risks analysis and allow us to check the behaviour of the estimates in finite samples. We report the squared bias (SB), variance (VAR), mean squared error (MSE), and the coverage probability (CP) for the estimated $(\tau, \gamma, \beta_{11}, \beta_{22})$.

The results for $\tau=0.5$ are reported in Table \ref{tab:res1}. The results for $\tau =0.1$ and $\tau=0.9$ are shown in Appendix I because they show the same main patterns as those for $\tau=0.5$. It can be seen in Table \ref{tab:res1} that the estimators have small SB and MSE for all considered sample sizes. These diminish further when the sample size increases. Most of the MSE comes from the variance whereas the SB are generally very small. The coverage probabilities are mostly equal to 0.95 or above. These results confirm that the partial likelihood estimator has nice finite and asymptotic properties. In addition, it is computationally convenient.

\begin{table}[htb]
	\centering
	\begin{adjustbox}{max width=\linewidth}
		\begin{threeparttable}
			\centering
			\caption{Simulation results. $\tau =0.5$.}
			\label{tab:res1}
			\begin{tabular}{lrrrrrrr}\hline\hline
     $n$ 	& 100	  &  200 		 & 400 	&		    & 100 		& 200 			& 400  \\\hline
   	  &  \multicolumn{3}{c}{{$\hat{\tau}$}} &	& \multicolumn{3}{c}{{$\hat{\gamma}$}}  \\\hline
SB    & 8.6e-04     & 1.0e-04		& 7.6e-06    &  & 1.1e-03	& 7.5e-05	  & 1.5e-05 	      \\
VAR   &	9.3e-02     & 3.7e-02 	& 7.1e-04       &  & 1.1e-02	& 5.1e-02	  & 1.1e-02 	   \\ 		
MSE		& 9.4e-02     & 3.7e-02		& 7.2e-04       &  & 1.1e-02 & 5.2e-02 	& 1.1e-02 	      \\
CP    & 0.99        & 0.99		  & 1.00       &  & 0.95	  & 0.95      & 0.96              \\\hline
      & \multicolumn{3}{c}{{$\hat{\beta}_{11}$}} &	&  \multicolumn{3}{c}{{$\hat{\beta}_{12}$}}  \\\hline
SB    & 3.4e-03	&  1.2e-04 	& 1.7e-05		  & & 2.6e-04 & 4.4e-04 	& 1.5e-05 			   \\
VAR  	& 9.0e-02 &  3.8e-02 	& 1.9e-02	    &	& 5.0e-02	& 2.4e-02 	& 1.1e-02 	 \\
MSE 	& 9.3e-02 &  3.8e-02 	& 1.90-02 	  & & 5.0e-02 & 2.4e-02	  & 1.1e-02	 \\
CP    & 0.97    &  0.96     & 0.95        &	& 0.94    & 0.95		  & 0.96        \\
\hline \hline
			\end{tabular}
		\end{threeparttable}	
	\end{adjustbox}
\end{table}

According to Corollary \ref{cor3}, a constant log hazard CSH model is inconsistent with the structural model for the competing risks model of Theorem \ref{ass1}. Despite this theoretical inconsistency, it is possible for a researcher to fit a Cox PH CSHs model as in (\ref{coxh}). In this case, $\alpha_j$ cannot be considered as a fixed parameter as it depends among other things on $z$ and its distribution (compare (\ref{comp1})). In consequence, the estimated $\alpha_j$ is sensitive in $z$ and its marginal distribution given the same structural parameters $(\theta, \gamma, \beta_{11}, \beta_{12})$. We illustrate this point with the hep of simulations where we draw samples from the same model except that the marginal distribution of $z$ is different. In particular, we consider estimation of $\alpha_j$ when the standard deviation of $z$ increases.

\begin{figure}[h]
     \centering
     \begin{subfigure}[b]{0.48\textwidth}
         \centering
         \caption{$\hat{\alpha}_1$}
         \includegraphics[height=8cm]{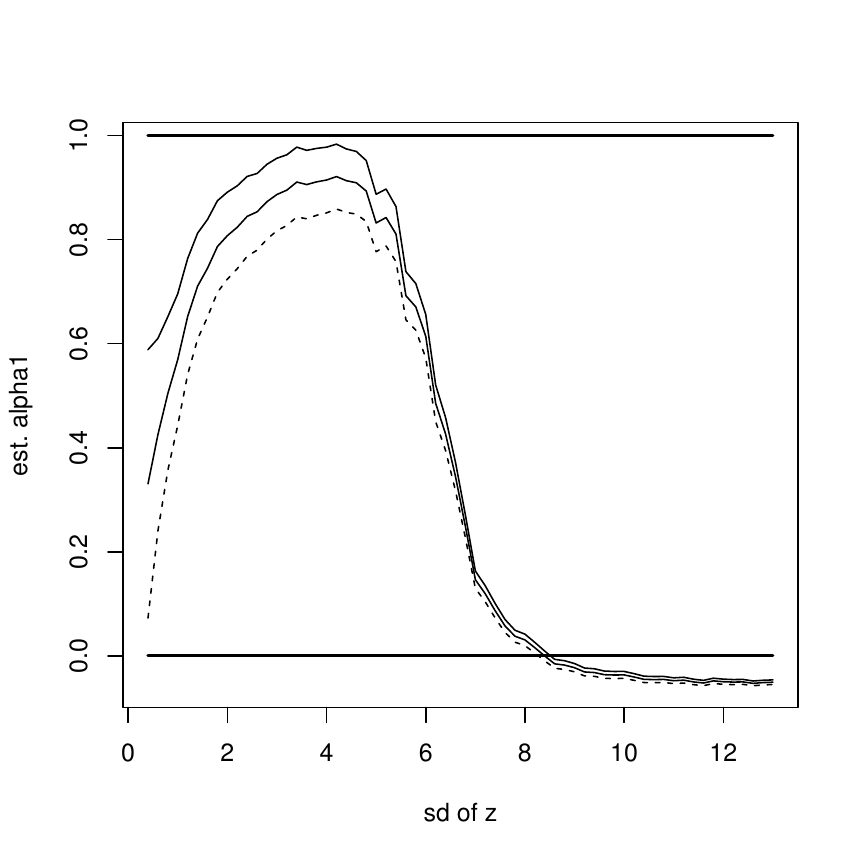}
         \label{figa1zsd}
     \end{subfigure}
     \hfill
		\begin{subfigure}[b]{0.48\textwidth
}
         \centering
         \caption{$\hat{\alpha}_2$}
         \includegraphics[height=8cm]{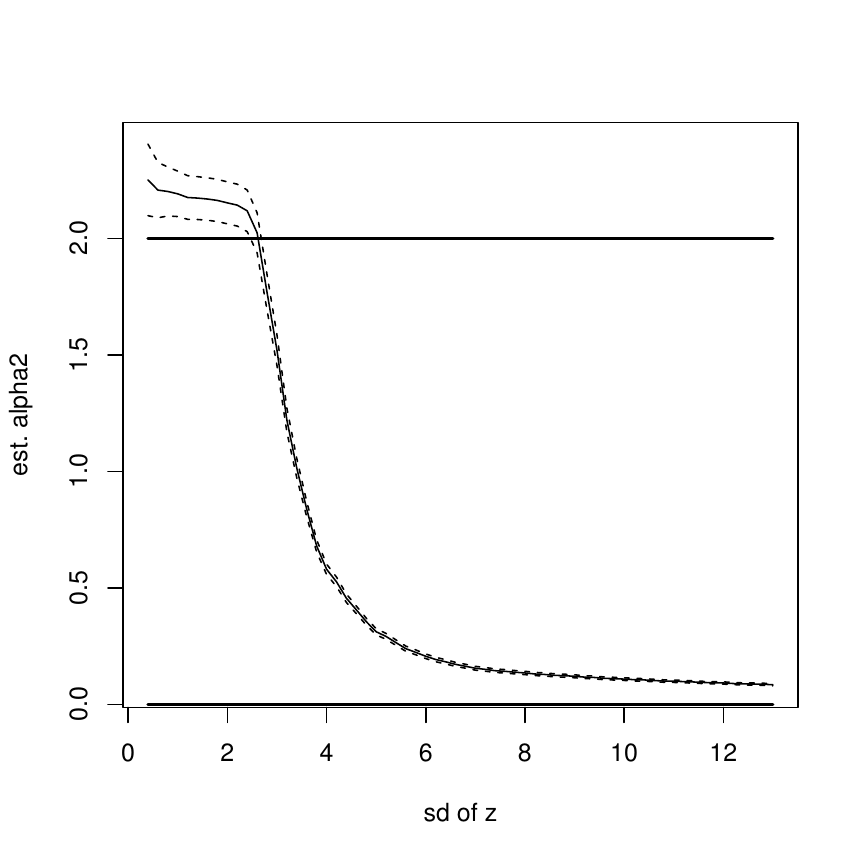}
         \label{figa2zsd}
     \end{subfigure}
					           \caption{Estimated $\alpha_j$ as a function of $\sigma_z$: Mean of estimated $\alpha_1$ and $\alpha_2$ with 5th and 95th percentile of the distribution of estimates. $n=5,000$, $\tau=0.5$.}
        \label{fig:azsd}
\end{figure}

Figure \ref{fig:azsd} shows how the estimated $\alpha_j$ changes with the standard deviation of $Z$ ($\sigma_z$) increases from 0.2 to 14. On a 0.2 grid on this range of $\sigma_z$ we draw 100 samples of size $n=5,000$ and display the mean of estimated $\alpha_j$ and the 5th and 95th percentile of their distribution. For comparison, we also plot $\beta_{1j}$ ($\beta_{11} =1$, $\beta_{12} =2$) of the DGP as a horizontal line. The estimated $\alpha_j$s are obviously not a constant, and they possess a nonlinear relationship with $\sigma_z$. The mean of the estimates of $\alpha_1$ ranges from -0.05 to 0.9, implying that the estimated log hazard ratio on the CSH ranges from $e^{-0.05} = 0.95$ to $e^{0.9} = 2.7$. Inference based on the PH CSH model with a constant log hazard ratio model can suggest a very strong effect of $Z$ in one data set when the s.d. of $Z$ is about five. In this case, it is estimated to increase the CSH by 2.7 times. In another data set from the same model just with a different s.d. of $z$ the estimated role of $Z$ can also be negligible or even slightly negative. For the latter case, $Z$ is estimated to reduce the hazard for risk 1 when the s.d. of $Z$ is around 9 or larger. Irrespective of the value of $\sigma_z$, the estimated effect of $Z$ on the CSH for risk 1 is weaker than the causal effect on the marginal hazard of risk 1. For risk 2, we also observe a strongly changing role of $Z$ when $\sigma_z$ increases. The mean of the estimated $\alpha_2$ has an even bigger range from $e^{0.1} = 1.1$ to $e^{2.2} = 9.0$. The estimated effect of $Z$ on the CSH on risk 2 is much smaller than the causal effect on the marginal hazard of risk 2, when $\sigma_z$ is greater than 3.

\begin{figure}[h]
     \centering
     \begin{subfigure}[b]{0.48\textwidth}
         \centering
         \caption{$\hat{\alpha}_1$}
         \includegraphics[height=8cm]{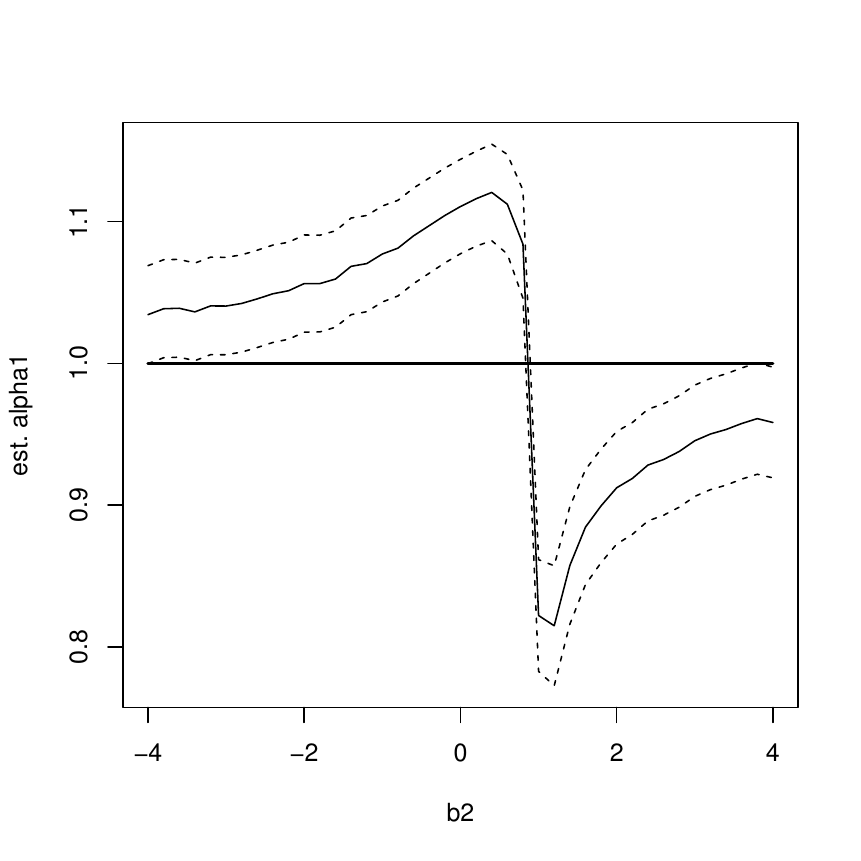}
         \label{figa1zsd}
     \end{subfigure}
     \hfill
		\begin{subfigure}[b]{0.48\textwidth}
         \centering
         \caption{$\hat{\alpha}_2$}
         \includegraphics[height=8cm]{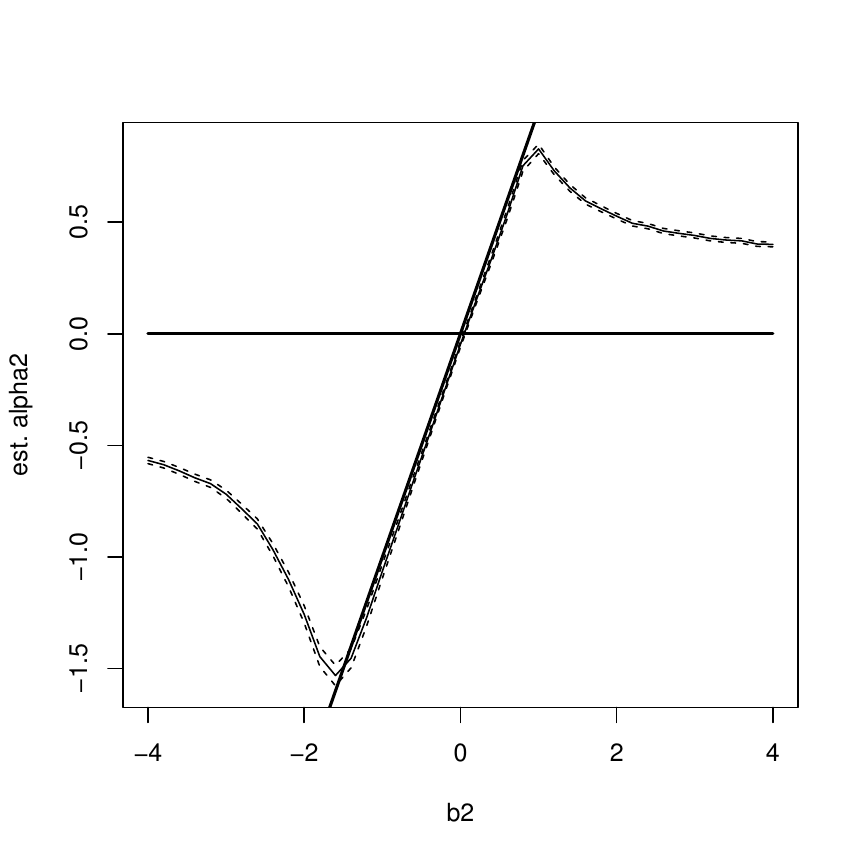}
         \label{figa2zsd}
     \end{subfigure}
					           \caption{Estimated $\alpha_j$ as a function of $\beta_{12}$: Mean of estimated $\alpha_1$ and $\alpha_2$ with 5th and 95th percentile of the distribution of estimates. $n=5,000$, $\tau=0.5$.}
        \label{fig:ab2}
\end{figure}

Next, we explore how the structural parameters affect the estimated $\alpha_j$. Figure \ref{fig:ab2} show once again the mean and the 5th and 95th percentile of the distribution of $\hat{\alpha}_j$ as a function of $\beta_{12}$. We consider the range from -4 to 4 with a grid distance of 0.2. As expected, the estimated $\alpha_1$ for risk 1 is affected by the covariate effect on risk 2 ($\beta_{12}$) even if the covariate effect on risk 1 $(\beta_{11})$ is constant. The estimated $\alpha_1$ behaves differently depending on the size of $\beta_{12}$ relative to $\beta_{11}$.  The estimated $\alpha_1>1$ when $\beta_{12} < \beta_{11} = 1$, whereas the estimated $\alpha_1<1$ when $\beta_{12}>\beta_{11} = 1$. Interestingly, the estimated $\alpha_1$ drops substantially when $\beta_{12}$ is close to $\beta_{11}$. The estimated log hazard ratio for $\alpha_1$ attains a maximum  of $e^{1.1} = 3.0$ when $\beta_{12}$ is about zero, and it has its minimum $e^{0.8} = 2.2$ when $\beta_{12}$ is greater than 1. Apart from this, there is no evidence of a tractable relationship between $\alpha_1$ and $\beta_{12}$. This confirms that the estimated $\alpha_1$ is hardly informative about the structural parameters that reflect the causal relationships. The effect of changing $\beta_{12}$ on $\alpha_2$ shows a completely different picture. The estimated $\alpha_2$ is very close to $\beta_2$ (the 45 degree line) when $\beta_2$ is between minus two and one. This linear relationship does not hold when $\beta_2$ is either more negative or more positive. The estimated log hazard ratio for $\alpha_2$ attains its maximum $e^{0.8} = 2.2$) when $\beta_{12}$ is about $1$, and its minimum $e^{-1.5} = 0.2$ when $\beta_{12}$ is about $-2$. Once again, it is impossible to infer anything about the $\beta_{12}$ when studying the estimated $\alpha_2$.

\begin{figure}[h]
     \centering
     \begin{subfigure}[b]{0.48\textwidth}
         \centering
         \caption{$\hat{\alpha}_1$}
         \includegraphics[height=8cm]{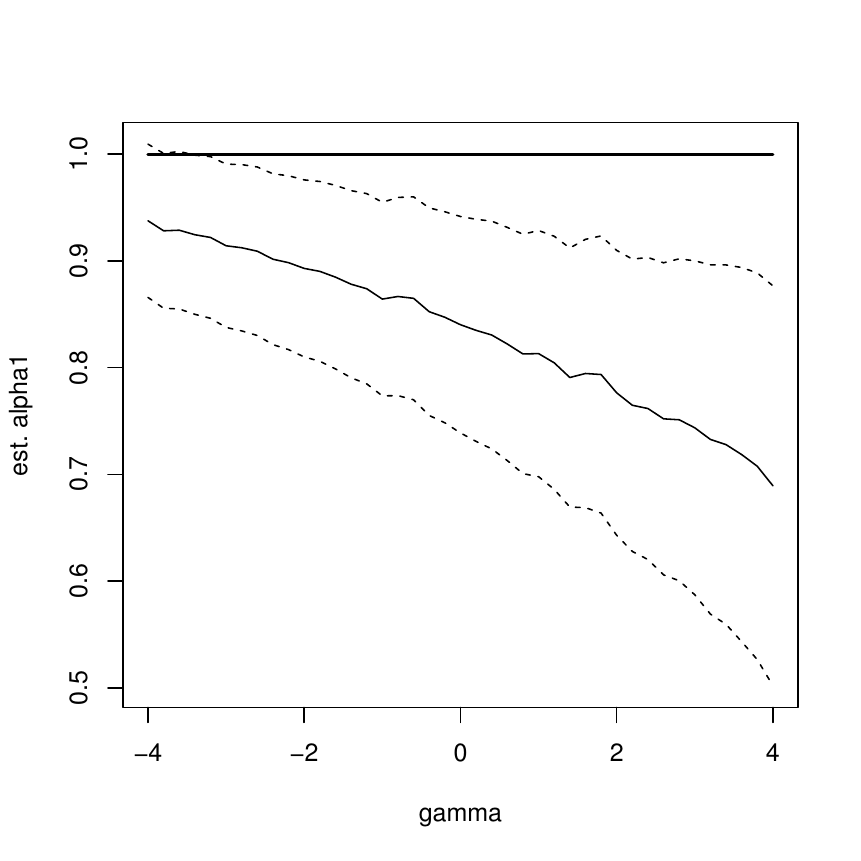}
         \label{figa1zsd}
     \end{subfigure}
     \hfill
		\begin{subfigure}[b]{0.48\textwidth}
         \centering
         \caption{$\hat{\alpha}_2$}
         \includegraphics[height=8cm]{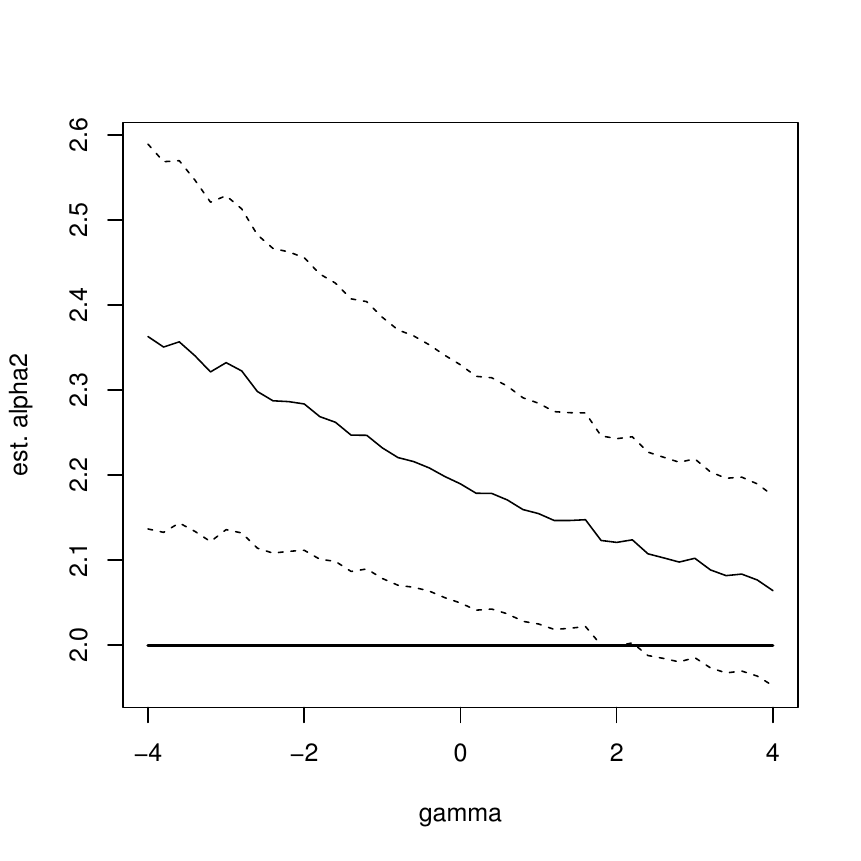}
         \label{figa2zsd}
     \end{subfigure}
					           \caption{Estimated $\alpha_j$ as a function of $\gamma$: Mean of estimated $\alpha_1$ and $\alpha_2$ with 5th and 95th percentile of the distribution of estimates. $n=5,000$, $\tau=0.5$.}
        \label{fig:agam}
\end{figure}

Lastly, we consider how $\alpha_j$ responds to a change in $\gamma$. The greater $\gamma$, the greater the baseline hazard for risk 2 compared to risk 1 which corresponds to an increase in incidences of risk 2 compared to risk 1.  It is noteworthy that the estimated $\alpha_1$ and $\alpha_2$ change with $\gamma$ despite that the causal effect of $Z$ on both marginal hazards is unchanged (i.e., $\beta_1$ and $\beta_2$ do not change). 

\section{Application \label{sec:appl}}

We illustrate the suggested methodology by using an economic data set on single job durations. Lo et al. (2024) describe these data in detail. It contains job tenures of employees that have one job. These can be terminated by two reasons. Risk 1 occurs when the worker quits the job to take a new job (job change), while risk 2 occurs then the worker accept an additional job in addition to the existing job (multiple jobs). It is interesting to distinguish these two risks, as holding a single job or holding multiple jobs involve different motivations and lead to different job search strategies (Hlouskova et al., 2017; Manchino and Mullins, 2019; Lal\'{e}, 2020; Lo, 2023). One should therefore expect the hazard rates for the two risks to be different. We include three covariates in the model, namely the current weekly working hours (divided by 168 = $24\times7$ hours), the current earned wage (measured in log), and a binary variable for male or female. The sample consists of 2,000 job durations. 63 percent of which are terminated by risk 1 and 37 percent are terminated by risk 2. The average job tenure is 59 weeks for risk 1 and 66 weeks for risk 2. It indicates that it takes a shorter time to change a job than the decision to take an additional job.

\begin{table}[htb]
	\centering
	\begin{adjustbox}{max width=\linewidth}
		\begin{threeparttable}
			\centering
			\caption{Estimation results using job duration data.}
			\label{t:res}
			\begin{tabular}{l c c c c  c c c c}\hline\hline
                &	\multicolumn{4}{c}{Structural model - $\beta$}   & \multicolumn{4}{c}{reduced-form model - $\alpha$}  \\ \hline
                & coef.  &  s.e. & t & h.r.                        & coef.  & s.e. & t & h.r. \\\hline
								  \multicolumn{7}{l}{Risk 1 - Changing jobs} \\
 								Weekly working hours   & 2.58    & 0.35  &7.38$^{**}$  & 13.2   &1.53  & 0.34 & 4.51$^{**}$     & 4.63  \\
								Log real hourly wage   & -0.43   & 0.04  &-11.7$^{**}$ & 0.65   &-0.39  & 0.04 & -9.71$^{**}$   & 0.68  \\
		            Female         & 0.04    & 0.05  & 0.89$ $     & 1.04   &-0.08  & 0.05 & -1.52$^{\dag}$     & 0.92  \\\hline
								 \multicolumn{7}{l}{Risk 2 - Taking a second job } \\
 	 						  Weekly working hours   & -1.84   & 0.50  &-3.70$^{**}$ & 0.16   &-5.29  & 0.43 & -12.4$^{**}$   & 0.01  \\
							  Log real hourly wage   & -0.23   & 0.05  &-4.55$^{**}$ & 0.80   &-0.13 & 0.05  & -2.37$^{*}$    & 0.88 \\
		            Female   			 & 0.12    & 0.06  & 2.11$^{*}$  & 1.13   & 0.16 & 0.07  & 2.44$^{*}$     & 1.18 \\\hline
								$\hat{\gamma}$				 &-0.95    & 0.41  &-2.32$^{*}$  & 0.39   &        &      &            &     \\
								$\hat{\theta}$  			 & 1.35    & 0.21  & 6.50$^{**}$ &        &        &      &            &    \\
\hline\hline
			\end{tabular}
\begin{tablenotes}
        \item Legend: $**$: 1\% significance level, $**$:  5\% significance level, $\dag:$  10\% significance level
    \end{tablenotes}
		\end{threeparttable}	
	\end{adjustbox}
\end{table}

We estimate the structural parameters ($\theta, \gamma, \beta$) using our proposed method, where $\theta$ is a copula parameter, $\gamma$ is the risk proportionality parameter, and $\beta$ is a vector of regression coefficients. For comparison, we also estimate the reduced-form parameters ($\alpha$) using the Cox PH CSHs model. The results are reported in Table \ref{t:res}. The estimated $\theta$ is 1.35 and is significant at the 1\% level. This implies a Kendall's $\tau$ of 0.26, i.e. there is moderate rank correlation. The estimated $\gamma$ is -0.95 and significant at the 5\% level. Using the relationship $\varsigma = \gamma/\theta$, the estimated $\varsigma$ is -0.70. It suggests that the marginal baseline hazard for risk 2 is half the marginal baseline hazard for risk 2 because $\lambda_{02}(t)/ \lambda_{01}(t) = e^{-0.70}= 0.50$. This explains why the job duration with end reason risk 2 is longer than that for risk 1.

Next, we discuss the estimated causal effect $\beta$. The structural model suggests that the hazard rate for job change (risk 1) increases significantly if the current job has a longer working hours ($\beta = 2.58$, hazard ratio = 13.2). Since we have normalised the working hours by the total available weekly hours 168, an increase in working hours by 168 raises the hazard rate by 13.2 times. As explained by Lo (2024), overworking (working too long) is a common phenomenon, but workers usually do not have much negotiating power to change their working hours in the present job. Finding a new job with fewer working hours is therefore a common strategy. It explains why we obtain a positive effect of working hours on the hazard rate of risk 1. In contrast, the effect of working hours on taking a second job (risk 2) is significantly negative. If the wage increased by 100\%, the hazard rate of finding a second job drops by 35\% (= 1-0.65). This result is also plausible, as underworking (working insufficient hours) is one of the major reasons for taking a second job, so that they can earn more income. The more working hours in the current job, the lower are the incentives to take an additional job. This reduces the hazard rate of risk 2. We can compare these estimated causal effects of working hours on the marginal hazards ($\beta$) with the estimated reduced-form effect on the CSH ($\alpha$). The $\alpha$ for both risks have the same sign as the corresponding $\beta$, but the sizes are rather different. The causal effect of working hours is greater on the hazard of risk 1 ($\beta = 2.58$) than for risk 2 ($\beta = 1.84$). The opposite holds for the reduced-form effects on risk 1 ($\alpha$ = 1.53) and risk 2 ($\alpha$ = 5.29). Although the reduced-form effect gives us a correct picture that more working hours result in a higher incidence of risk 1 ($\Pr(\delta =1)$ increases) and a shortened job duration, it fails to explain the underlying reasons -- how working hours affect risks 1 and 2 differently. In fact, if we solely relied on the estimated $\alpha$, we may be tempted to draw a conclusion that the negative causal effect of working hours on risk 2 is of greater magnitude than the positive causal effect on risk 1. This is however not the case.

We briefly discuss the second and the third covariate in the following. The causal effect of earned wage acts negatively on both risks ($\beta = -0.43$ on risk 1 and $\beta = -0.23$ on risk 2). Increasing the current wage by one percent reduces the marginal hazard rates by 35\% (= 1-0.65) and 20\% (=1-0.80) for risks 1 and 2, respectively. These results are economically sensible. People tend to stay at the current job when the current wage is higher. Also, there is a smaller incentive to take a second job when the current wage is higher. The reduced-form effects on the CSHs ($\alpha$) have the same direction as the causal effects, and they have similar sizes as well. Lastly, the covariate female is insignificant for risk 1. Females are not more or less likely to change their jobs than males. In contrast, the estimated reduced-form effect is negative and marginally significant as the p-value for a one-tail test is 0.064. It illustrates again that the structural model may lead the researcher to come to a different conclusion than the structural model for the causal effects. $\beta$ for risk 2 is significantly positive. Females are more likely than men to take a second job, with a marginal hazard rate higher by 13\%. It is comprehensible as females are more likely than men to hold a part-time job or irregular jobs, which should facilitate taking a second job. The estimated reduced-form effect for risk 2 is also significantly positive, although it has a larger size.

To summarise, the estimated reduced-form effects using the CSH model are generally different from the estimated causal effects using the structural model. Even though the two risks are only found to be mildly dependent in our application with the Kendall's $\tau$ of 0.26, the results demonstrate that it is impossible to infer anything from the reduced-form model for the causal effects of the structural model.

\newpage
\section*{Appendix I - Tables}

\begin{table}[htb]
	\centering
	\begin{adjustbox}{max width=\linewidth}
		\begin{threeparttable}
			\centering
			\caption{Simulation results. $\tau =0.1$}
			\label{tab:res1}
			\begin{tabular}{lrrrrrrr}\hline\hline
   $n$ 	& 100	  &  200 		 & 400 	&		    & 100 		& 200 			& 400  \\
   	  &  \multicolumn{3}{c}{{$\hat{\tau}$}} &	& \multicolumn{3}{c}{{$\hat{\gamma}$}}  \\\hline
SB    & 4.1e-03 & 1.1e-03		& 9.4e-06    &  & 3.5e-04	& 9.7e-05	  & 1.7e-03 	      \\
VAR   &	0.53    & 0.21 	    & 8.6e-02       &  & 7.7e-02	& 3.7e-02	  & 1.8e-02 	   \\ 		
MSE		& 0.53    & 0.21		  & 8.6e-02       &  & 7.7e-02 & 3.70e-02 	& 1.9e-02 	      \\
CP    & 0.89    & 0.93		  & 0.97       &  & 0.95	  &  0.95     & 0.95              \\
      & \multicolumn{3}{c}{{$\hat{\beta}_{11}$}} &	&  \multicolumn{3}{c}{{$\hat{\beta}_{12}$}}  \\\hline
SB    & 3.2e-03	&  1.1e-03 	& 6.0e-05		  & & 1.3e-05 & 3.8e-05 	& 4.9e-05 			   \\
VAR  	& 0.14    &  6.1e-02 	& 2.5e-02	    &	& 6.8e-02	& 3.1e-02 	& 1.4e-02 	 \\
MSE 	& 0.14    &  6.20-02 	& 2.50-02 	  & & 6.8e-02 & 3.1e-02	  & 1.4e-02	 \\
CP    & 0.94    &  0.96     & 0.97        &	& 0.95    & 0.95		  & 0.96        \\
\hline \hline
			\end{tabular}
		\end{threeparttable}	
	\end{adjustbox}
\end{table}

\begin{table}[htb]
	\centering
	\begin{adjustbox}{max width=\linewidth}
		\begin{threeparttable}
			\centering
			\caption{Simulation results. $\tau =0.9$}
			\label{tab:res1}
			\begin{tabular}{lrrrrrrr}\hline\hline
   $n$ 	& 100	  &  200 		 & 400 	&		      & 100 		& 200 			& 400  \\
   	  &  \multicolumn{3}{c}{{$\hat{\tau}$}} &	& \multicolumn{3}{c}{{$\hat{\gamma}$}}  \\\hline
SB    & 1.1e-05 & 2.5e-07		& 7.62e-06      &  & 1.8e-02	& 1.9e-02	 & 2.8e-03 	      \\
VAR   &	3.7e-03 & 1.8e-03 	& 7.1e-04       &  & 2.2      & 0.43	   & 0.13  	   \\ 		
MSE		& 3.7e-03 & 1.8e-03		& 7.2e-04       &  & 2.2      & 0.45 	   & 0.14  	      \\
CP    & 1.00    & 1.0		    & 1.00          &  & 0.96	    & 0.97     & 0.96              \\
      & \multicolumn{3}{c}{{$\hat{\beta}_{11}$}} &	&  \multicolumn{3}{c}{{$\hat{\beta}_{12}$}}  \\\hline
SB    & 2.2e-03	   &  1.9e-03 	& 1.2e-04		  & & 7.8e-04 & 1.6e-04 	& 2.0e-06 			   \\
VAR  	& 7.5e-02    &  3.0e-02 	& 1.4e-02	    &	& 4.4e-02	& 2.0e-02 	& 9.0e-03 	 \\
MSE 	& 7.7e-02    &  3.20-02 	& 1.40-02 	  & & 4.5e-02 & 2.0e-02	  & 9.0e-03	 \\
CP    & 0.93       &  0.95      & 0.95        &	& 0.94    & 0.94		  & 0.95        \\
\hline \hline
			\end{tabular}
		\end{threeparttable}	
	\end{adjustbox}
\end{table}

\newpage
\section*{Appendix II - Violation of the PH assumption for Clayton copula}

The Clayton copula is $C(u,v|\theta) = [u^{-\theta} + v^{-\theta} -1 ]^{-1/\theta}$ with $\theta \in [-1,\infty)\ \{0\}$. The two risks are independent for $\theta =0$. Their correlation increases with $\theta$. The implied CSH $h^{s}_j(t|z)  = f^{s}_j(t|z)/S(t|z)$ for $j=1,2$ is then
\begin{eqnarray}
h^{s}_j(t|z) = S(t|z)^{\theta} S_j(t|z)^{-\theta}\lambda_j(t|z). \label{hstrclayton}
\end{eqnarray}
\nin It is clear that $h^{s}_j(t|z)$ does not satisfy the PH assumption in (\ref{coxh}) as it is generally not separable in $t$ and $z$. It is obvious that separability of the marginal hazards, i.e. separability of $\Lambda_{j0}(t)$ and $\phi_j(z)$ in $S_j(t|z) = \exp[-\Lambda_{j0}(t)\phi_j(z)]$, does not imply separability of $h^s_j(t|z)$ in (\ref{hstrclayton}). This is only the case for $\theta =0$ when the CSHs are identical to the marginal hazards, i.e., $h^{s}_j(t|z) = \lambda_j(t|z)$ for $j=1,2$. In this case, the PH assumption on $\lambda_j(t|z)$ implies the same for $h^{s}_j(t|z)$. In other words, the PH assumption for CSHs model is only consistent with a structural model when there are independent risks.

\newpage
\section*{Appendix III - R scripts}

\scriptsize
\begin{singlespace}
\begin{verbatim}



# original data frame
d<-data.frame(t,del,z)

# Restructuring the data
d1<-d[which(d[,2]==1),]
d2<-d[which(d[,2]==2),]
z11<-cbind(d1[,3:ncol(y1)],matrix(0,nrow=nrow(d1), ncol =ncol(d1)-2))
z12<-cbind(matrix(0,nrow=nrow(d1), ncol =ncol(d1)-2),y1[,3:ncol(d1)])
z21<-cbind(d2[,3:ncol(y2)],matrix(0,nrow=nrow(d2), ncol =ncol(d2)-2))
z22<-cbind(matrix(0,nrow=nrow(d2), ncol =ncol(d2)-2),y2[,3:ncol(d2)])
y11<-cbind(d1[,1],rep(1,nrow(d1)),d11)
y12<-cbind(d1[,1],rep(0,nrow(d1)),d12)
y21<-cbind(d2[,1],rep(0,nrow(d2)),d21)
y22<-cbind(d2[,1],rep(1,nrow(d2)),d22)

# Restructured data frame
Y<- rbind(y11,y12,y21,y22)
Z <- as.matrix(cbind(Y$z1,Y$z2,Y$z3,Y$z4))
rt <- as.vector(Y$rt)
n.obs <- sum(rt)
x <- as.vector(Y$x)
zz1<-as.vector(Y$z1)
zz2<-as.vector(Y$z2)
zz3<-as.vector(Y$z3)
zz4<-as.vector(Y$z4)

# Define risk set
risk.set <- function(t) which(x >= t)

# Create risk set
rs <- apply(as.matrix(x[as.logical(rt)]), 1, risk.set)

# Partial likelihood function
log.parlik3 <- function(beta){
  a1<-0         # constraint
  b1<-beta[1]
  a2<-beta[2]
  b2<-beta[3]
  theta <-beta[4]

  A<- (1+exp(a2)*exp((zz2+zz4) *(b2-b1)*theta))^(1/theta-1)
  lnA <- log(A)
  b2s <- b1*(1-theta) +b2*theta
  status <- as.vector(as.logical(rt))
  Xbeta <- as.vector(zz1*a1+zz2*b1+zz3*a2+zz4*b2s)
  lpl1 <- sum(Xbeta[status]+lnA[status])
  temp <- vector(   )
  for(i in 1:n.obs)  temp[i] <- log(sum(exp(Xbeta[rs[[i]]]+lnA[rs[[i]]])))
  lpl2 <- sum(temp)
  return(-lpl1 + lpl2)
}

# Run optimization
OPT <- optim(c(2,1,1,1), log.parlik3, control = list(maxit = 1000), hessian=TRUE)
\end{verbatim}
\end{singlespace}
\end{document}